\documentstyle{article}

\title{\huge Axion detection by ring lasers\thanks{
hep-ph/9502256, submitted to Phys.\,Lett.\,B}}
\author{L. Cooper\thanks
{Present address: Department of Physics, Theoretical Physics,
1 Keble Road, Oxford~OX1\,3NP. E-mail: leith@thphys.ox.ac.uk}
\ and G.E. Stedman\thanks
{E-mail: g.stedman@csc.canterbury.ac.nz}\\ \\
{\em Department of Physics and Astronomy\/}\\
{\em University of Canterbury\/}\\
{\em Private Bag 4800, Christchurch\/}\\
{\em New Zealand.\/}}
\date{\today}

\begin{document}

\maketitle

\begin{abstract}
A ring-laser experiment, similar to the Canterbury ring laser, to detect
axion- and {\sc qed}-induced vacuum birefringence is proposed. It uses a
slowly modulated magnetic field and a novel polarization geometry. Both
axion coupling and vacuum birefringence would modulate the Sagnac beat
frequency.  A null result could place sensitive bounds on the axion mass and
on two-photon coupling.
\end{abstract}

\section*{}
The axion was originally introduced \cite{pecceiET1977a} to
explain the absence of $CP$ violation in the strong interaction. Axions also
arise in supersymmetric and superstring theories and are candidates for dark
matter in the universe. Laboratory experiments and a host of astrophysical
arguments \cite{turner1990a} constrain the axion mass to between
$10^{-6}$\thinspace eV and $10^{-3}$\thinspace eV. A number of optical
experiments have searched for such axions which couple to two photons either
directly or via ``triangle diagrams'' of charged particles.
Sikivie \cite{sikivie1983a} suggested that relic axions could be
detected through their conversion to photons in a resonant cavity bathed in
a strong magnetic field. Experiments of this sort, which rely on the
assumption that axions form a significant component of the galactic halo,
have yet to confirm the presence of axions \cite{dePanfilisET1987a}. Attempts
to produce axions and detect their reconversion into photons, the so-called
``shining-light-through-walls'' experiment, \cite{anselm1985a} have, so far,
given null results \cite{cameronET1993a}.
Maiani et al. \cite{maianiET1986a,raffeltET1988a} have suggested that
axions induce small
perturbations in the polarization state of a laser beam propagating in a
magnetic field. The absence of optical rotation and ellipticity in the
transmitted beam has been used to place limits on the axion mass and
coupling to two photons \cite{cameronET1993a,semertzidisET1990a}. We adapt
the ideas of Maiani et al. to propose and examine the feasibility for a
ring-laser experiment suitable for detecting axion- and {\sc qed}-induced
vacuum birefringence. We suggest that a ring-laser experiment could play a
significant role in the search for axions.

Ring lasers have primarily been used to detect absolute rotation
\cite{andersonET1994a}. A ring laser is, however, much more than a
gyroscope \cite{stedmanET1993a}. A rotational frequency splitting
results from the nonreciprocal change $\Delta \!L$ in the optical path
lengths for the
counterpropagating beams. Viewed from the rotating system, this may be
regarded as the consequence of a difference between the effective refractive
indices for the two beams. Conversely, any birefringence resulting in a
difference $\Delta \!n$ in the refractive indices will induce a beat
frequency,
\begin{equation}
\Delta \!f=\frac{f\Delta \!L}L=\frac{f\Delta \!n\,l}L\,,  \label{eqsagnac2}
\end{equation}
where $l$ is the length of the cavity over which the birefringence
occurs, $L$ is the ring perimeter and $f$ the optical frequency. In
this sense, a
ring laser acts as a sensitive differential refractometer
\cite{bilgerET1993a}. More generally, it has recently been shown that the beat
frequency in a ring laser originates in either time-reversal or parity
violating effects, depending on the polarization geometry
\cite{bilgerET1994a}. The He\thinspace Ne ring laser system installed at
Cashmere, Christchurch, New Zealand (henceforth called the Canterbury ring
laser), has attained \cite{bilgerET1994b} a frequency resolution
$\delta f\sim 140\,$ nHz, or 1 part in $3\times 10^{21}$ of the optical
frequency $f=474$ \thinspace THz. From Eq.\thinspace (\ref{eqsagnac2}), this
in turn implies a measurement capability $\Delta n=\Delta L/l\sim 10^{-20}$,
for a nonreciprocal contribution to the refractive index in a sample of
length $l=10$\thinspace cm with $L=3.4771$\thinspace m. At this level
of precision, a ring-laser
detection of axions and even {\sc qed} photon-photon coupling becomes
feasible. The potential for a precision ring-laser to place sensitive bounds
on field-induced birefringence of the vacuum has been recognized
\cite{stedmanET1993a,stedmanET1987a}.  However, the specific details for such
experiments have not previously been discussed.

In a square planar-ring, the out-of-plane ($s$) component of electric field
is usually excited rather than the in-plane ($p$) component since, according
to the Fresnel reflection coefficients, the $p$ reflection is inevitably
more lossy for oblique incidence. The proposed polarization geometry is
illustrated in Fig.\thinspace 1. Two Faraday cells, whose optic axes are
diametrically opposed in the direction of beam propagation, are placed at
each end of one of the legs of the ring. The polarization vector $\vec s
=(1,0)$ for the incident beam is rotated successively into $(a,b)=R(\theta
)(1,0)$ and $R(-\theta )(a,b)=\vec s$ by the Faraday cell at each end of
the leg. Accordingly, $s$ polarization is maintained in the ring as a whole,
thus ensuring high mirror reflectivity. Within the leg, the
counterpropagating beams will have mutually orthogonal polarizations for a
Faraday rotation angle $\theta =45^{\circ }$. Such a polarization geometry
could feasibly be engaged by a pair of terbium-gallium-garnet ({\sc tgg})
Faraday rotators (Litton-Airtron Industries), cut at Brewster's angle, which
have at least 99.9\thinspace \% transmission at 633.0\thinspace nm.
Such a total loss, of the order of 1000\thinspace ppm, would degrade
the cavity quality factor implied above by a factor of up to 100, and
requires fuller design work. We propose at least the use of
Brewster-cut end faces to help ensure that the fraction of these
losses contributing to backscatter and so frequency pulling remains at
an acceptable level (say $\leq 50$\thinspace ppm).

The coupling of axions to two photons via the triangle anomaly is described
by the effective lagrangian density
\begin{equation}
{\cal L}=\frac 1M\,(\vec E \cdot \vec B_{\mbox{\tiny ext}})\,a,  \label{eqeb}
\end{equation}
where $\vec E$ is the electric field vector of the laser beam propagating
through an applied (static) magnetic field $\vec B_{\mbox{\tiny ext}}$
and $a$  is
the pseudoscalar axion field. The inverse coupling $M=g_{\gamma \gamma
a}^{-1}$ has dimensions of energy and a natural interpretation as a
symmetry-breaking scale. If the axion mass $m_a$ is less than the photon
energy $\omega $, axions can be produced through the Primakoff effect
\cite{primakoff1951a}  according to the graph in Fig.\thinspace
2\thinspace (a).
In view of Eq.\thinspace (\ref{eqeb}), the component of $\vec E$ parallel
to $\vec B_{\mbox{\tiny ext}}$ is attenuated while the orthogonal component is
unaffected. Production of virtual axions is shown in Fig.\thinspace
2\thinspace (b), whereby the parallel component of the incident laser beam
oscillates to a massive axion for part of its travel, and so is retarded
with respect to the orthogonal component. This phase shift would be
converted to a frequency shift in an active ring laser, as the corresponding
wavelength stretches or shrinks to accommodate the new round-trip optical
path length. This effect can take place even if $m_a>\omega $.

When the mixing of the photon and axion fields is weak
\cite{raffeltET1988a}, the phase shift and attenuation of $\vec
E_{\parallel}$ are
\begin{eqnarray}
\delta (l) &=&2\left( \frac{\omega B_{\mbox{\tiny
ext}}}{Mm_a^2}\right) ^2\sin^2\left( \frac{m_a^2l}{4\omega }\right)
\,,  \label{eqdeltal} \\
\phi _a(l) &=&\left( \frac{\omega B_{\mbox{\tiny
ext}}}{Mm_a^2}\right)^2\left(
\frac{m_a^2l}{2\omega }-\sin \frac{m_a^2l}{2\omega }\right) \,,
\label{eqphil}
\end{eqnarray}
where $l$ is the length of the magnetic field region. Up to a factor of $%
L=Nl $, these results agree with Eqs.\thinspace (42) and\thinspace (43) of
Ref. \cite{raffeltET1988a}. This factor arises in multipass
cavity ellipsometry \cite{cameronET1993a,semertzidisET1990a} where
the laser beam is multiply reflected back and forth $N$ times through the
magnetic field so as to accumulate a larger effect. The measured rotation, $%
\epsilon =\delta /2$, and ellipticity, $\psi =\phi /2$, are cumulative upon
reflection so that for $N$ reflections, the values of $\epsilon $ and $\psi $
given by Eqs.\thinspace (\ref{eqdeltal}) and\thinspace (\ref{eqphil}) are
increased by a factor of $N$. In an active ring-laser, however, the
frequency shift is determined solely by the nonreciprocal change in the
optical path length for one complete circuit of the ring. In this sense, the
proposed ring-laser experiment is a single-pass experiment with $N=1$ and $%
L=l$.

For photon energies $\omega <2m_e$, the birefringence due to the {\sc qed}
vacuum-polarization graph of Fig.\thinspace 2\thinspace (c) is
\cite{adler1971a}:
\begin{equation}
\phi _{\mbox{\tiny QED}}(l)=\frac{2\alpha ^2B_{\mbox{\tiny ext}}^2}{15m_e^4}\,
\omega l\,.
\label{eqqedphi}
\end{equation}
where $m_e$ is the electron mass and $\alpha =e^2/4\pi \approx 1/137$ is the
fine structure constant. Vacuum birefringence is particularly interesting as
a test of higher-order {\sc qed} perturbation theory which has not, as yet,
been measured directly with real incident photons \cite{iacopiniET1979a}.

If $\vec B_{\mbox{\tiny ext}}$ is modulated at low frequency, then the
selective
attenuation and retardation of one of the ring-laser beams will induce {\sc %
am} and {\sc fm} sideband peaks to the earth's rotation-induced Sagnac beat
frequency, with peak amplitude and phase deviations $\delta $ and $\phi $,
respectively. Analyzing the amplitude, phase and harmonic content of the
sideband spectrum should thus be sufficient to determine the mass, coupling
and parity of the axion. The details of sideband analysis in a precision
ring laser has been discussed elsewhere \cite{bilgerET1994b}. For {\sc fm} at
a frequency $f_{\mbox{\tiny m}}$, we may write the instantaneous
ring-laser beat frequency as
\[
f=f_{\mbox{\tiny c}}-\Delta \!f\sin \omega _{\mbox{\tiny m}}t\,,
\]
where $\Delta \!f$ is the peak frequency deviation from the Sagnac carrier
frequency $f_{\mbox{\tiny c}}$. The resultant {\sc fm} signal can be
expanded as
\begin{eqnarray*}
V(t)\Big|_{\mbox{\tiny FM}} &=& \Re\left\{ V_{\mbox{\tiny c}}\exp i(\omega
_{\mbox{\tiny c}}t+\beta \cos \omega _{\mbox{\tiny m}}t)\right\}  \\
&=&\Re\left\{ V_{\mbox{\tiny c}}(\exp i\omega _{\mbox{\tiny
c}}t)\Big(J_0(\beta
)+2\sum_{n=1}^\infty i^nJ_n(\beta )\cos (n\omega _{\mbox{\tiny
m}}t)\Big) \right\}
\,,
\end{eqnarray*}
where the modulation index $\beta =\Delta \!f/f_{\mbox{\tiny m}}$ is
the  amplitude
(in radians) of the phase excursion induced by the frequency modulation and
\begin{equation}
J_n(\beta )=\left( \frac \beta 2\right) ^n\sum_{m=0}^\infty (-1)^m\frac{%
(\beta /2)^{2m}}{m!\,(m+n)!}  \label{eqseries}
\end{equation}
is the $n$th-order Bessel function. This result tells us that with
single-frequency dithering, there will be an infinite series of sideband
satellites above and below the earth line, separated in frequency precisely
by $f_{\mbox{\tiny m}}$, with amplitudes controlled by $J_n(\beta
)$. When $\beta
\ll 1$, Eq.\thinspace (\ref{eqseries}) may be expanded as
\[
J_0(\beta )\approx 1-\frac {\beta^2}{4},\ \ \ J_1(\beta )\approx
\frac{\,\beta }{%
16}(8-\beta ^2)\ \ \ \mbox{and}\ \ \ J_2(\beta )\approx \frac{\,\,\beta ^2}{%
96}(12-\beta ^2)\,.
\]
Interpreting the ring-laser detection voltage as quantifying a light beam
intensity rather than an amplitude, the heights of the lowest-order
sidebands are $N$ decibels below the carrier, where
\begin{equation}
N=10\log _{10}\!\left( \frac{J_0(\beta )}{J_1(\beta )}\right) \approx 10\log
_{10}\!\left( \frac 2\beta \left(1-\frac 18 \beta^2\right)\right)
\approx 3-10\log_{10}\!\beta \,.  \label{eqsn}
\end{equation}
The carrier amplitude itself is reduced by $-10\log _{10}\!J_0(\beta
)\approx \beta ^2\,$dB. Note that contrary to standard engineering practice,
we follow Ref. \cite{bilgerET1994b} and define the
decibel as $10\log _{10}\!V$, rather than $20\log _{10}\!V$, since the
photomultiplier signal $V$ is already proportional to the light intensity,
rather than its amplitude. From Eq.\thinspace (\ref{eqsn}), a sideband which
is 10$n$\thinspace dB below the carrier translates into a modulation
index $\beta =2\times 10^{-n}$ and so to a peak frequency deviation
$\Delta\!f=2\times 10^{-n}f_{\mbox{\tiny m}}$.

Fig.\thinspace 3 plots the signal-to-noise ({\sc s}/{\sc n}) ratios thus
required to detect a given frequency shift $\Delta \!f$ for $\mu$Hz,
mHz and Hz modulation frequencies. The shaded region indicates
the present sensitivity of the Canterbury ring laser: for a modulation
frequency of 1\thinspace Hz, a peak frequency shift of 1\thinspace $\mu $Hz
would yield sidebands (displaced 1\thinspace Hz either side of the Sagnac
carrier) which were resolvable above the noise. From Eq.\thinspace (\ref
{eqsagnac2}), this translates into a sensitivity for detecting a
nonreciprocal contribution $\Delta \!n=7.3\times 10^{-20}$ to the refractive
index in a sample of length $l=10\,$cm. This in turn implies a measurement
capability
\begin{equation}
\Delta \phi =\frac{2\pi f\,\Delta \!n\,l}c=7.3\times 10^{-14}
\label{eqphasemod}
\end{equation}
for the axion-induced phase shift between the counterpropagating
beams.

Conversely, an observed {\em absence\/} of {\sc am} and {\sc fm}
sidebands would, by virtue of Eqs.\thinspace (\ref{eqdeltal}) and
(\ref{eqphil}),  place upper bounds on the axion-photon coupling
$g_{\gamma \gamma a} = M^{-1}$. These bounds are illustrated in
Fig.\thinspace 4 as a function of axion mass, assuming a ring-laser
sensitivity as given in Eq.\thinspace (\ref{eqphasemod}), for a
1\thinspace T magnetic field applied across a 10\thinspace cm section
of the beam paths. Various experimental and  theoretical bounds on the
axion have been summarized in
Ref. \cite{cameronET1993a} and are also shown in Fig.\thinspace 4.
For $m_a<10^{-3}$\thinspace eV, we see that the optical-rotation bound of
the Brookhaven-Fermilab-Rochester ({\sc bfr}) multipass cavity experiment is
an order-of-magnitude better than that which could be achieved by a
ring-laser experiment with $\mu $Hz frequency resolution. This is despite
the fact that the ring-laser experiment could detect phase shifts $\phi \sim
10^{-13}$\thinspace rad, some $10^4$-times better than the corresponding
measurements made in the {\sc bfr} experiment. This is because the magnitude
of the induced effects depend critically upon the strength of the magnetic
field and the optical path length $L=Nl$ through the field region. However,
for relatively large axion masses, $m_a>2\times 10^{-3}$\thinspace eV, the
ring laser could surpasses the corresponding {\sc bfr} ellipticity bound.
This is because the ellipticity measurements were three times less sensitive
than the optical rotation measurements due to larger systematic effects and
also a more pronounced effect of random motion of the beam on the
ellipsometer optics \cite{cameronET1993a}.

With improved sensitivity, the ring laser could probe higher effective
energies, thus placing tighter bounds on the axion mass and coupling to two
photons. In order to detect {\sc qed} vacuum birefringence,
Eq.\thinspace (\ref{eqqedphi})  implies that a beat frequency
resolution of 54\thinspace pHz
would be required. Fig.\thinspace 3 shows that for modulation of $\vec
B_{\mbox{\tiny ext}}$  at 1\thinspace Hz, the {\sc qed}-induced sidebands would
lie 110\thinspace dB below the Sagnac carrier. Although the present
signal-to-noise ratio is approximately 60--90\thinspace dB on a hertz-wide
scale, a synchronous filter could feasibly extract such a signal from the
noise. We have not explored the limits of phase-sensitive or similar such
{\sc ac} methods of detection. Indeed, the possibility of a ring-laser
detection of {\sc qed} vacuum birefringence provides strong incentive for
further research in this direction.

To summarize, astrophysical and laboratory constraints leave but one window
open for the axion mass: $10^{-6}$\thinspace eV to $10^{-3}$\thinspace eV.
The proposed ring-laser experiment could search for axions in this mass
range which couple to two photons upto an effective energy of
$10^5$\thinspace GeV, assuming microhertz resolution of the ring-laser beat
frequency. The {\sc bfr} experiment \cite{cameronET1993a},
however, has already ruled out the existence of such axions with couplings
below $3\times 10^6$\thinspace GeV. Moreover, calculations based upon
stellar evolution \cite{turner1990a} limit $M>3\times 10^9\,$GeV
for axions of mass $10^{-6}$\thinspace eV to $10^{-3}$\thinspace eV.
Nevertheless, a ring-laser experiment could still provide independent
insight into the question of axion parameters.

We emphasize that the above estimates have been based upon the level of
precision (1 $\mu $Hz) achieved by the Canterbury ring laser in early 1994,
and which has already been exceeded by a factor 6. With improved frequency
resolution, assisted by phase-sensitive methods of detection, the bounds
placed by a ring-laser experiment could surpass those bounds already placed
by the {\sc bfr} experiment. Construction of a second-generation ring-laser
at Canterbury has been initiated, which should afford up to a 100-fold
improvement in resolution \cite{andersonET1994a}. At this level of precision,
we believe that a ring-laser experiment to place sensitive new bounds on the
axion, by searching for axions in the mass range $10^{-6}$\thinspace
eV to $10^{-3}$\thinspace eV which couple to two photons up to an
effective energy of $2\times 10^7$\thinspace GeV, would be a realistic
goal. This goal would be further motivated by the possibility of a
ring-laser detection of {\sc qed} vacuum birefringence. We suggest
that the time is opportune to commence an experiment on the above
lines in parallel with the theoretical investigation of
external-field-induced effects in ring lasers.

\section*{Acknowledgements}
{\sc lc} acknowledges partial support from a University of Canterbury
Masters' Scholarship, and advice from Scott Griffen, Litton-Airtron
Industries re specifications of {\sc tgg} Faraday rotators.

\newpage\

Captions to figures:

Fig.\thinspace 1. Ring-laser polarization geometry for detecting
axions and {\sc qed} vacuum birefringence. Attenuation and retardation
of the beam polarized parallel to the external magnetic field
modulates the ring laser output signal.

Fig.\thinspace 2. (a) Photons propagating in a transverse magnetic field
producing axions through the Primakoff effect. (b) Virtual axion
production leading to vacuum birefringence. (c) {\sc qed} vacuum
birefringence via $\gamma $-$\gamma $ scattering.

Fig.\thinspace 3. Canterbury ring-laser {\sc fm}
sensitivity. $\Delta\! f$ is the peak frequency deviation in Hz. The
shaded region indicates ring-laser {\sc s}/{\sc n}  ratios at
present. The dashed line indicates the {\sc s}/{\sc n} ratios required
to detect {\sc qed} vacuum birefringence for $\mu $Hz, mHz and Hz
modulation of $\vec B_{\mbox{\tiny ext}}$.

Fig.\thinspace 4. Limits on the axion mass $m_a$ and coupling $g_{\gamma
\gamma a}$ to two photons \cite{cameronET1993a}. The heavy dashed
line indicates the sensitivity of the proposed ring-laser experiment,
conservatively assuming a $\mu $Hz level frequency resolution. The shaded
region indicates the results from the {\sc bfr} experiment.


\begin{thebibliography}{99}

\bibitem{pecceiET1977a}
R.D. Peccei and H.R. Quinn, Phys. Rev. Lett. 38 (1977) 1440;\\
R.D. Peccei and H.R. Quinn, Phys. Rev. D 16 (1977) 1791;\\
S. Weinberg, Phys. Rev. Lett. 40 (1978) 223;\\
F. Wilczek, Phys. Rev. Lett. 40 (1978) 279.

\bibitem{turner1990a}
M.S. Turner, Phys. Rep. 197 (1990) 67;\\
G.G. Raffelt, Phys. Rep. 198 (1990) 1;\\
M.A. Bershady, M.T. Ressell and M.S. Turner,
Phys. Rev. Lett. 66 (1991) 1398.

\bibitem{sikivie1983a}
P. Sikivie Phys. Rev. Lett. 51 (1983) 1415, Erratum 52 (1984) 695;\\
P. Sikivie Phys. Rev. D 32 (1985) 2988.

\bibitem{dePanfilisET1987a}
S. DePanfilis, A.C. Melissinos, B.E. Moskowitz, J.T. Rogers,
Y.K. Semertzidis W.U. Wuensch, H.J. Halama, A.G. Prodell, W.B. Fowler
and F.A. Nezrick, Phys. Rev. Lett. 59 (1987) 839;\\
W.U. Wuensch, S. DePanfilis-Wuensch, Y.K. Semertzidis, J.T. Rogers,
A.C. Melissinos, H.J. Halama, B.E. Moskowitz, A.G. Prodell,
W.B. Fowler and F.A. Nezrick Phys. Rev. D 40 (1989) 3153;\\
C. Hagmann, P. Sikivie, N.S. Sullivan and D.B. Tanner,
Phys. Rev. D 42 (1990) 1297;\\
D.M. Lazarus,  G.C. Smith, R. Cameron, A.C. Melissinos, G. Ruoso,
Y.K. Semertzidis and F.A. Nezrick, Phys. Rev. Lett. 69 (1992) 2333.

\bibitem{anselm1985a}
A.A. Ansel'm, Sov. J. Nucl. Phys. 42 (1985) 936;\\
K. van Bibber, N.R. Dagdeviren, S.E. Koonin, A.K. Kerman and
H.N. Nelson Phys. Rev. Lett. 59 (1987) 759.

\bibitem{cameronET1993a}
R. Cameron, G. Cantatore, A.C. Melissinos, G. Ruoso, Y. Semertzidis,
H.J. Halama, D.M. Lazarus, A.G. Prodell, F. Nezrick, C. Rizzo and
E. Zavattini, Phys. Rev. D 47 (1993) 3707.

\bibitem{maianiET1986a}
L. Maiani, R. Petronzio and E. Zavattini,
Phys. Lett. B 175 (1986) 359.

\bibitem{raffeltET1988a}
G. Raffelt and L. Stodolsky, Phys. Rev. D 37 (1988) 1237.

\bibitem{semertzidisET1990a}
Y. Semertzidis, R. Cameron, G. Cantatore, A.C. Melissinos, J. Rogers,
H.J. Halama, A.G. Prodell, F. Nezrick, C. Rizzo and E. Zavattini,
Phys. Rev. Lett. 64 (1990) 2988.

\bibitem{andersonET1994a}
R. Anderson, H.R. Bilger and G.E. Stedman,
Am. J. Phys. 62 (1994) 975.

\bibitem{stedmanET1993a}
G.E. Stedman, H.R. Bilger, Z. Li, M.P. Poulton, C.H. Rowe,
I. Vetharaniam and P.V. Wells, Aust. J. Phys. 46 (1993) 87.

\bibitem{bilgerET1993a}
H.R. Bilger, G.E. Stedman, M.P. Poulton, C.H. Rowe, Z. Li and
P.V. Wells,
IEEE Trans. Instrum. Meas. 42 (1993) 407, Erratum 43 (1994) 102.

\bibitem{bilgerET1994a}
G.E. Stedman, M.T. Johnsson, Z. Li, C.H. Rowe and H.R. Bilger,
T violation and microhertz resolution in
a ring laser (1994) Opt. Lett. to be published.

\bibitem{bilgerET1994b}
G.E. Stedman, Z. Li and H.R. Bilger,
Sideband analysis and seismic detection in a precision ring laser
(1994) Appl. Opt. submitted for publication.

\bibitem{stedmanET1987a}
G.E. Stedman and H.R. Bilger, Phys. Lett. A 122 (1987) 289.

\bibitem{primakoff1951a}
H. Primakoff, Phys. Rev. 81 (1951) 899.

\bibitem{adler1971a}
S.L. Adler Ann. Phys. 67 (1971) 599.

\bibitem{iacopiniET1979a}
E. Iacopini and E. Zavattini, Phys. Lett. B 85 (1979) 151;\\
E. Iacopini, B. Smith, G. Stefanini and E. Zavattini,
Nuovo Cim. B 61 (1981) 21;\\
G. Cantatore, F. Della Valle, E. Milotti, L. Dabrowski and C. Rizzo,
Phys. Lett. B 265 (1991) 418;\\
Wei-Tou Ni, Kimio Tsubono, Norikatsu Mio, Kazumichi Narihara, Shen-Che
Chen, Sun-Kun King and Sheau-Shi Pan,
Mod. Phys. Lett. A 6 (1991) 3671.

\end{thebibliography}
\end{document}